\documentclass{PoS}

\usepackage{psfig}

\title{LOFAR, LEAP and beyond: Using next generation telescopes for pulsar astrophysics}

\ShortTitle{LOFAR, LEAP and beyond}

\author{\speaker{Michael Kramer}\thanks{also: Jodrell Bank Centre for
    Astrophysics, The University of Manchester}\\
        MPI f\"ur Radioastronomie, Auf dem H\"ugel 69, 53121 Bonn, Germany\\
        E-mail: \email{mkramer@mpifr.de}}

\author{Ben Stappers\\
        Jodrell Bank Centre for Astrophysics, The University of
        Manchster, Manchester M13 9PL, UK\\
        E-mail: \email{Ben.Stappers@manchester.ac.uk}}

      \abstract{Radio astronomy has benefited greatly from advances in
        technology and will continue to do so in the future. In fact,
        we are experiencing a revolution in the way radio astronomy is
        conducted as our instruments allow us now to directly
        ``digitize'' our photons. This has enormous consequences,
        since we can greatly benefit from the continuing advances in
        digital electronics, telecommunication and computing.  The
        results are dramatic increase in observable bandwidths, FoVs,
        frequency coverage and collecting area.  The global efforts
        will culminate in the construction of the SKA as the world's
        largest and most powerful telescope. On the way projects like
        LOFAR, LEAP and others will revolutionize many areas of
        astrophysics and fundamental physics. Observations of pulsars
        will play a central role in these scientific endeavours. We
        briefly summarize here some recent scientific developments
        that help us in defining our expectations for the the new
        generation of radio telescopes for pulsar astrophysics.  }

\FullConference{ISKAF2010 Science Meeting - ISKAF2010\\
		June 10-14, 2010\\
		Assen, the Netherlands}

\begin{document}

\section{Introduction}

Pulsar astrophysics has an impressive track record of making
fundamental discoveries in a wide range of physics and astrophysics,
covering the parameter space from the smallest scales (i.e. doing
solid state physics under extreme conditions) to the largest scales
(e.g. by exploring the nature of gravity). Apart from discovering the
existence of neutron stars and their nature as gigantic quantum
mechanical objects \cite{hbp+68}, further discoveries include, for
instance, the first evidence for the existence of gravitational waves
\cite{tay94b}, the discovery of the first extrasolar planets
\cite{wf92} and the existence of very fast spinning objects
challenging proposed equation-of-states \cite{bkh+82}. Amazingly, in
the 40$+$ years since the serendipitous discovery of pulsars, the rate
of discovery has not slowed, driven by new technology and instruments
that allows us to expand the accessible parameter space by improving
the time and frequency resolution and observing bandwidth. However,
the greatest leap in technological advance is yet to come with the new
generation of radio telescopes. Facilities like LOFAR and the SKA will
provide us with much larger collecting area, larger fractional
bandwidth, larger fields-of-views and multi-beam capabilities.  This
combination will provide a massive increase in sensitivity combined
with the opportunity to monitor many sources with excellent
cadence. Based on current and past experience, we have very high
expectations of what we are going to find. Indeed, we have every
reason to believe that the previous amazing discoveries were just the
prelude to what we will be able to do in the future. This contribution
will try to glance into the future by providing some examples of
excellent recent discoveries or developments and what can be expected
when we will use the new facilities or re-use the existing ones in a
novel fashion like in the LEAP project.

\section{Finding new and exciting sources}

Previous experience has proven that finding a large number of new
pulsars will inevitably lead to the discovery of rare objects which
push our understanding of their formation or which can be used as
unique laboratories for fundamental physics. Currently, we know of
about 2000 radio pulsars of which about half were discovered in a
single survey, i.e.~the Parkes Multibeam Survey for Pulsars
\cite{mlc+01}. The most recent jump in the discovery of the
fast-spinning type of millisecond pulsars (MSPs) was achieved by
performing radio searches in unidentified gamma-ray point sources as
seen with the FERMI space-telescope \cite{rs10}. A total of 19 MSPs
were discovered so far, providing a ``fast-track'' method to identify
these most useful sources. On-going surveys which employ
high-sensitivity (like P-ALFA at the Arecibo telescope \cite{cfl+06})
or high-time and -frequency resolution (like the HTRU survey with the
Parkes and Effelsberg telescopes \cite{kjv+10}) are finding
interesting new pulsars. Indeed, the High-Time-Resolution Universe
(HTRU) survey will provide an all-sky inventory of the variable sky,
already discovering lots of new sources in a previously
well-searched area. So far, the HTRU survey has already discovered
about 40 slowly rotating pulsars and about 10 MSPs (including bright
sources, eclipsing systems and planetary companions - stay
tuned!). Perhaps, the most unexpected find of the HTRU survey so far
was the first ever discovery of a magnetar in a blind radio survey
\cite{lbb+10}. The source PSR J1622$-$4950 has a 4.3-s period and with
an inferred value of $B=3 \times 10^{14}$ G, the highest magnetic field
for any radio-discovered neutron star. Detections at X-rays show an
X-ray luminosity that is a third of the spin-down luminosity,
confirming that this magnetar was discovered in an X-ray quiescent
phase. Moreover there is no record of previous X-ray outbursts either.
Yet, it shows all the radio properties of a radio-transient
magnetar: strong variation in flux density and pulse shape, a flat
spectrum and a very high degree of polarisation. Archival radio
observations reveal that the source was radio active in the past, but
that it was missed in previous observations.

The discovery of transient phenomena like the new radio magnetar
underlines the need for repeated surveys of the whole sky. The new
facilities like LOFAR and SKA with their large field-of-view will
provide superb opportunities to undertake many such surveys in short
time intervals. That is even already true for shallow surveys, for
instance, with the Australian SKA Pathfinder (ASKAP) and the
South-African MeerKAT. In any case, a combination of sensitivity and
field-of-view will revolutionize our knowledge of the Galactic pulsar
population. LOFAR will find the local population of neutron stars and
will also be sensitive to those radio pulsars, which are too weak at
the usual high search frequencies due to the generally steep pulsar
flux density spectrum \cite{svk+07}.  This local census will be
complemented by the SKA and its Galactic census of pulsars, which will
essentially include every pulsar that is beaming towards Earth. The
20,000 to 30,000 pulsars to be discovered should join the about
1000 MSPs, about 100 relativistic binaries and eventually the rare
objects like pulsar-black hole systems \cite{kbc+04,sks+09}.

\begin{figure}
\centerline{\psfig{file=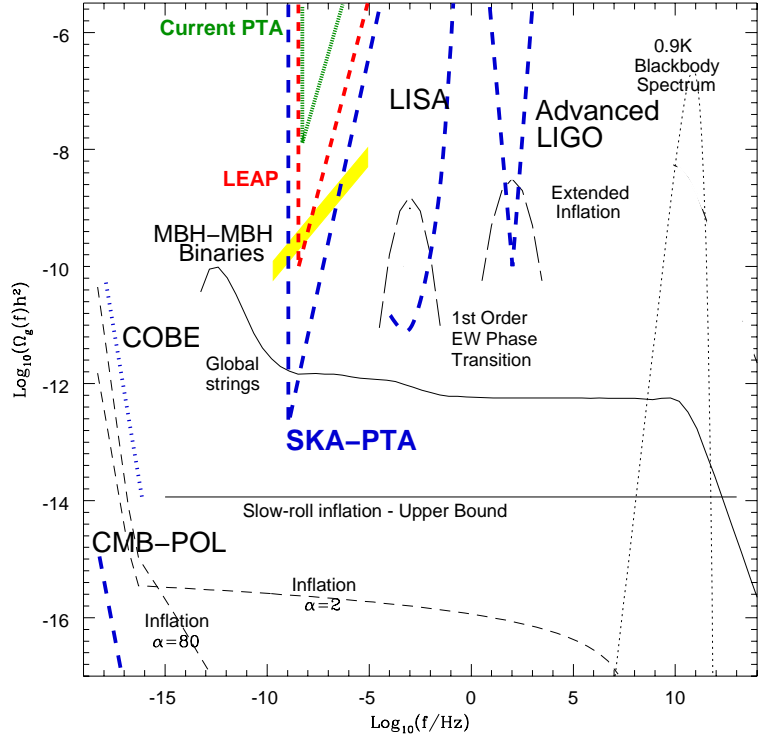,width=10cm}}
\caption{Summary of the potential cosmological sources 
of a stochastic gravitational wave (GW) background overlaid with
bounds from COBE, current Pulsar Timing Array (PTA) experiments and
the goals of 
CMB polarization experiments, LISA and Advanced LIGO. LEAP will improve on
the current best PTA limits by more than two orders of magnitude,
enabling the detection of a GW background caused by the merger of
massive black holes (MBHs) in early galaxy formation. The amplitude
depends on the MBH mass function and merger rate, so that uncertainty
is indicated by the size of the shaded area. LEAP is the next logical
step towards a PTA realized with the SKA which will improve on the
current sensitivity by about four orders of magnitude.
\label{fig:spectrum}}
\end{figure}

\section{Monitoring, watching and triggering}

The radio sky reveals variable and transient phenomena which are
relevant for a huge range of (astro-)physical questions. We have
already demonstrated above that repeated monitoring can reveal new,
previously unseen objects which are important for population studies
or the connections between radio and high-energy processes. Often
radio observations can trigger searches and studies in other observational
windows. Perhaps one of the most exciting such possibilities is the
triggering of gravitational wave searches for neutron star oscillations
after the detection of a pulsar glitch.

A pulsar glitch is a sudden increase in spin frequency of the neutron
star caused by an internal reconfiguration of the star's interior
structure (e.g.~\cite{ls04}). The relaxation of the glitch gives
information about the super-fluid interior of the pulsar and enables
us to do neutron star seismology. It is expected that such an event
may also cause the neutron star to oscillate, which for certain
vibration modes, should result in the emission of gravitational waves
(e.g.~\cite{vm08}). Knowing the exact moment of the glitch through
dense radio monitoring, gravitational wave data can be searched
accordingly. The statistical information obtained from such radio
monitoring with the new multi-beam capable telescopes will be used to
study the glitch mechanism and to answer questions as to whether a
possible bimodal distribution of the glitch sizes means that two
different types of glitch processes are acting (e.g.~\cite{esp10}).

Radio monitoring with good cadence is also essential to study the
recently discovered phenomenon that radio pulsars appear to be able to
change the structure of current flow in the magnetosphere in an
seemingly instantaneous way. This newly recognised class of
``intermittent pulsars'' \cite{klo+06}, seem to be active for a period
of time before switching off completely for a further period of time,
where the change in observed radio output is correlated with a change
in pulsar spin-down rate. In the case of PSR B1931+24, if the pulsar
is emitting radio waves, the spin-down rate is about 50\% larger than
when the pulsar is off. This faster spin-down rate is caused by an
extra torque that is given by the electric current of the plasma that
also creates the radio emission. If some or all of the plasma is
absent, the radio emission is missing together with the additional
torque component, so that the spin-down is slower. Very recently, Lyne
et al.~\cite{lhk+10} realized that intermittent pulsars are actually
the extreme form of a more common phenomenon, in which the
restructuring of the plasma currents does not necessarily lead to a
complete shutdown in the radio emission, but can be observed as
changes between distinct pulse shapes. Lyne et al.~showed that
particular profiles are indeed correlated with specific values for the
spin-down rates, confirming the previous picture. The times when the
switch in the magnetospheric structure occurs are for some pulsars
quasi-periodic but in general difficult to predict. The resulting
change between typically two spin-down rates leads to seemingly random
timing residuals that have been in the past classified as ``timing
noise''. The observations by Lyne et al.~therefore simultaneously
connect the phenomenon of timing noise to that of intermittent pulsars
and that of ``moding'' and ``nulling'' (see e.g.~\cite{lk05}).  An
interesting aspect is the possibility of determining the exact times
of the switch between magnetospheric changes by precisely measuring
the pulse shapes with high-sensitivity, high-cadence monitoring with
LOFAR or the SKA. In this case, the changes in spin-down rate can be
taken into account and the pulsar clock can be ``corrected''. This
would offer the opportunity to use not only the MSPs for timing
experiments, but to utilize also the 20,000 to 30,000 normal pulsars
that will be discovered in a Galactic census described above.  Even
though the precision will not be as high as for MSPs, the large number
of pulsars may help to detect, for instance, gravitational waves.

\section{Gravitational Wave detection}

The observed orbital decay in binary pulsars detected via precision
timing experiments so far offers the only evidence for the existence
of gravitational wave (GW) emission. Intensive efforts are therefore
on-going world-wide to make a direct detection of gravitational waves
that pass over the Earth. Ground-based detectors like GEO600 or LIGO
use massive mirrors, the relative distance of which are measured by a
laser interferometer set-up, while the future space-based LISA
detector uses formation flying of three test-masses that are housed in
satellites.  The change of the space-time metric around the Earth also
influences the arrival times of pulsar signals measured at the
telescope, so that high-precision MSP timing can also potentially
directly detect GWs. Because pulsar timing requires the observations
of a pulsar for a full Earth orbit before the relative position
between pulsar, Solar System Barycentre and Earth can be precisely
determined, only GWs with periods of more than a year can usually be
detected. In order to determine possible uncertainties in the used
atomic clocks, planetary ephemerides used, and also since GWs are expected
to produce a characteristic quadrupole signature on the sky, several
pulsars are needed to make a detection. The sensitivity of such a
``Pulsar Timing Array'' (PTA) increases with the number of pulsars and
should be able to detect pulsars in the nHz regime, hence below the
frequencies of LIGO ($\sim$kHz and higher) and LISA ($\sim\mu$Hz)
(see Fig.~\ref{fig:spectrum}).

A number of PTA experiments are ongoing, namely in Australia, Europe
and North America (see \cite{haa+10} for a summary). The currently
derived upper limits on a stochastic GW background
(e.g.~\cite{jhv+07,fvb+10}) are very close to the theoretical
expectation for a signal that originates from binary super-massive
black holes expected from the hierarchical galaxy evolution model
\cite{svc08,sv10}. It seems that ``simply a bit of extra sensitivity
is needed'' to make a first detection. This is the motivation for the
Large European Array for Pulsars (LEAP) project in Europe. It aims to
phase-coherently connect Europe's largest radio telescopes to form an
Arecibo-sized dish that can observe a large number of MSPs with high
sensitivity enabling high precision pulsar timing. LEAP is part of the
European Pulsar Timing Array (EPTA) and also acts as a test-bed for
SKA technology \cite{fvb+10}.

Demonstrating the power of PTA experiments, Champion et
al. \cite{ceh+10} recently used data of PTA observations to determine
the mass of the Jovian system independently of the space-craft data
obtained by fly-bys. Here, the idea is that an incorrectly known
planet mass will result in an incorrect model of the location of the
Solar System Barycentre (SSB) relative to the Earth. However, the SSB
is the reference point for pulsar arrival time measurements, so that a
mismatch between assumed and actual position would lead to a periodic
signal in the pulsar data with the period being that of the planet
with the ill-measured mass. This measurement technique is sensitive to
a mass difference of two hundred thousand million million tonnes --
just 0.003\% of the mass of the Earth, and one ten-millionth of
Jupiter's mass.

\begin{figure}
\centerline{\psfig{file=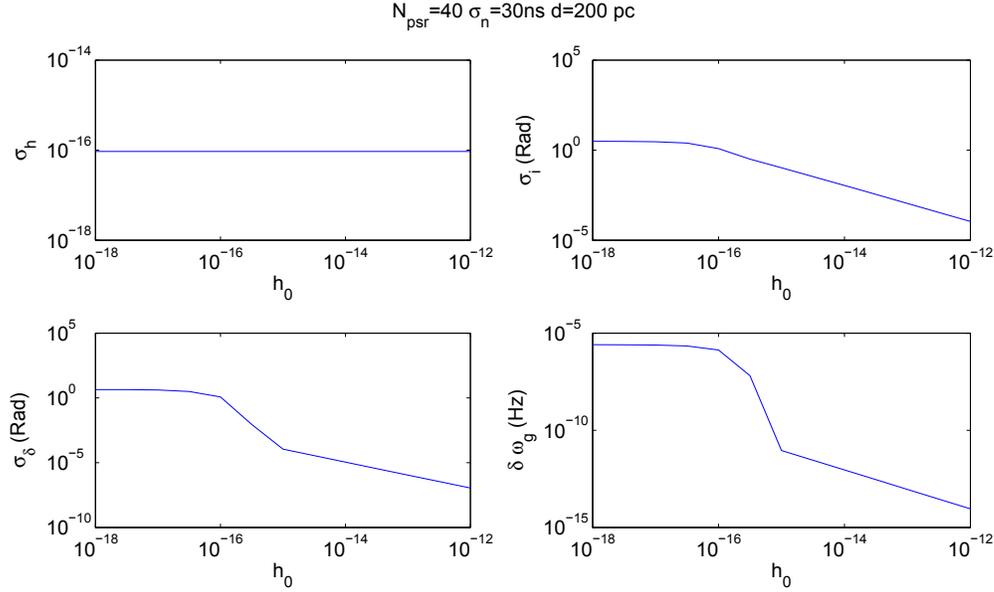,width=15cm}}
\caption{Results of computations for the detection of a single GW 
  source with the help of a PTA consisting of 40 pulsars with an average timing
  precision of 300 ns at a typical distance of 200 pc \cite{lwk+10}. 
{\em top left)}  Error in measuring the characteristic strain amplitude
of a single GW source for a variety of signal strengths.
{\em top right)} corresponding error in orbital inclination measurement,
{\em bottom left)} positional error on the sky and 
{\em bottom right)} error in determining the gravitational wave
  frequency.
See Lee et al. for details.
\label{fig:single}}
\end{figure}

If LEAP or other experiments do not detect GWs in the next few years,
a first detection is virtually guaranteed with the more sensitive
Phase I of the SKA.  But the science that can eventually be done with
the full SKA goes far beyond simple GW detection -- a whole realm of
astronomy and fundamental physics studies will become possible. For
instance, it will be possible to study the properties of the graviton,
namely its spin (i.e.~polarisation properties of GWs) and its mass
(note that in general relativity the graviton is massless) \cite{ljp08,ljp+10}.  This is
achieved by measuring the degree of correlation in the arrival time
variation of pairs of pulsars separated by a certain angle on the sky.
A positive correlation is expected for pulsars in the same direction
or 180 deg apart on the sky, while pulsars separated by 90 deg should
be anti-correlated. The exact shape of this correlation curve
obviously depends on the GW polarisation properties \cite{ljp08} but
also on the mass of the graviton \cite{ljp+10}. The latter becomes clear
when we consider that a non-zero mass leads to a dispersion relation
and a cut-off frequency $\omega_{\rm cut}=m_{\rm g}c^2/\hbar$, below
which a propagation is not possible anymore, affecting the degree of
correlation possible between two pulsars. With a 90\% probability,
massless gravitons can be distinguished from gravitons heavier than
$3\times 10^{-22}$\,eV (Compton wavelength $\lambda_{\rm g}=4.1
\times 10^{12}$ km), if bi-weekly observation of 60 pulsars are
performed for 5 years with pulsar RMS timing accuracy of 100\,ns. If
60 pulsars are observed for 10 years with the same accuracy, the
detectable graviton mass is reduced to $5\times 10^{-23}$\,eV
($\lambda_{\rm g}=2.5 \times 10^{13}$ km) \cite{ljp+10}.

In addition to detecting a {\em background} of GW emission, the
probability of detecting a {\em single} GW source increases from a few
percent now to well above 95\% with the full SKA. We can, for instance,
expect to find the signal of a single super-massive black hole
binary. Considering the case when the orbit is effectively not
evolving over the observing span, we can show that, by using
information provided by the ``pulsar term'' (i.e.~the retarded effect
of the GW acting on the pulsar's surrounding space time), we can
achieve a rather astounding source localization. For a GW with an
amplitude exceeding $10^{-16}$ and PTA observations of 40 pulsars with
weekly timing to 30 ns, precision one can measure the GW source position to an
accuracy of better than$\sim1$ arcmin (Fig.~\ref{fig:single}, \cite{lwk+10}). With such an error
circle, an identification of the GW source in the electromagnetic
spectrum should be easily feasible. We note that in order to achieve
such a result, a precise distance measurement to the pulsars is
needed, which in turn can then be improved further during the fitting
process that determines the orbital parameters of the GW
source. Fortunately, the SKA will be a superb telescope to do
astrometry with pulsars.

\section{Astrometry}

\begin{figure}
\centerline{
\begin{tabular}{cc}
\psfig{file=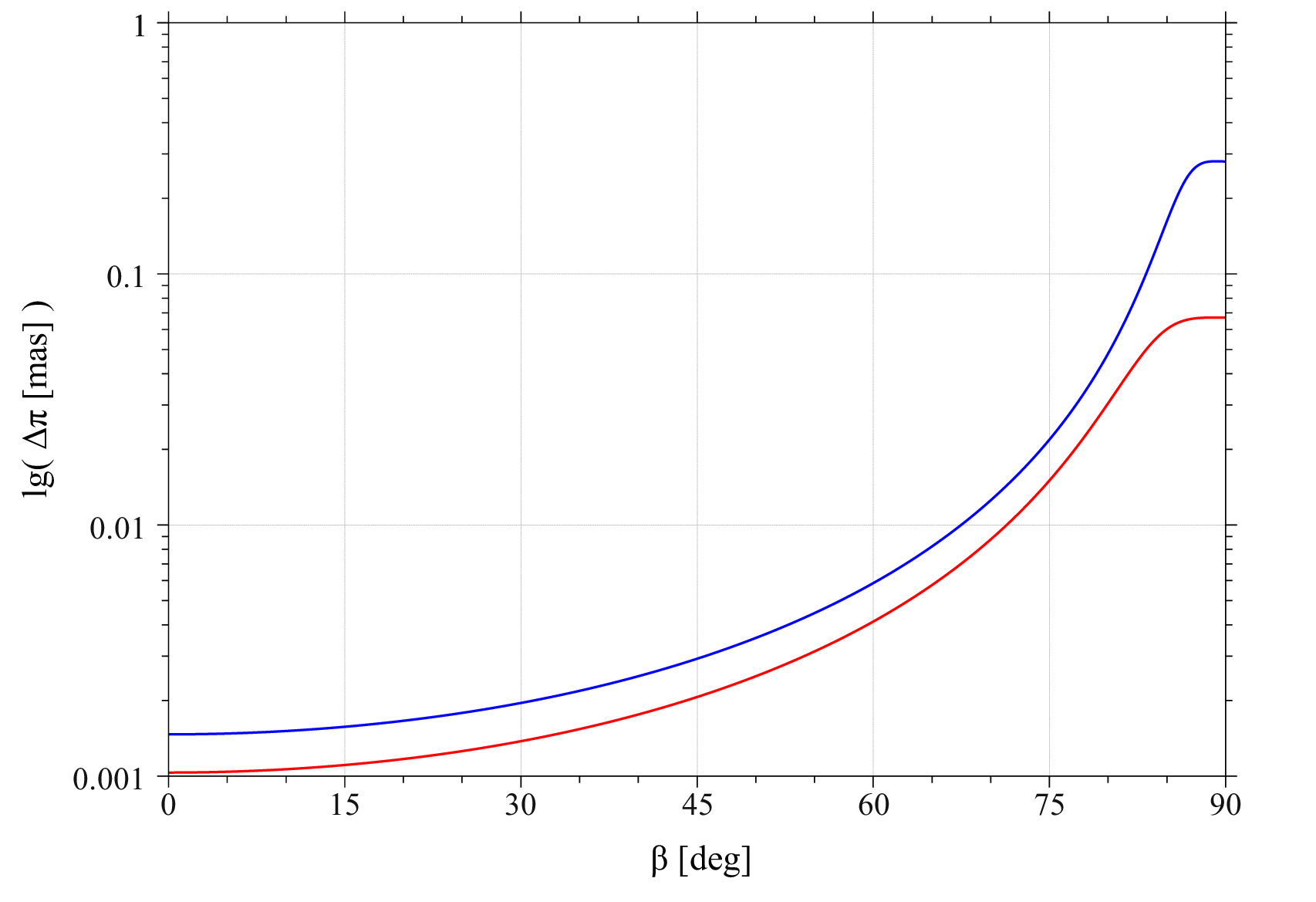,width=7cm} &
\psfig{file=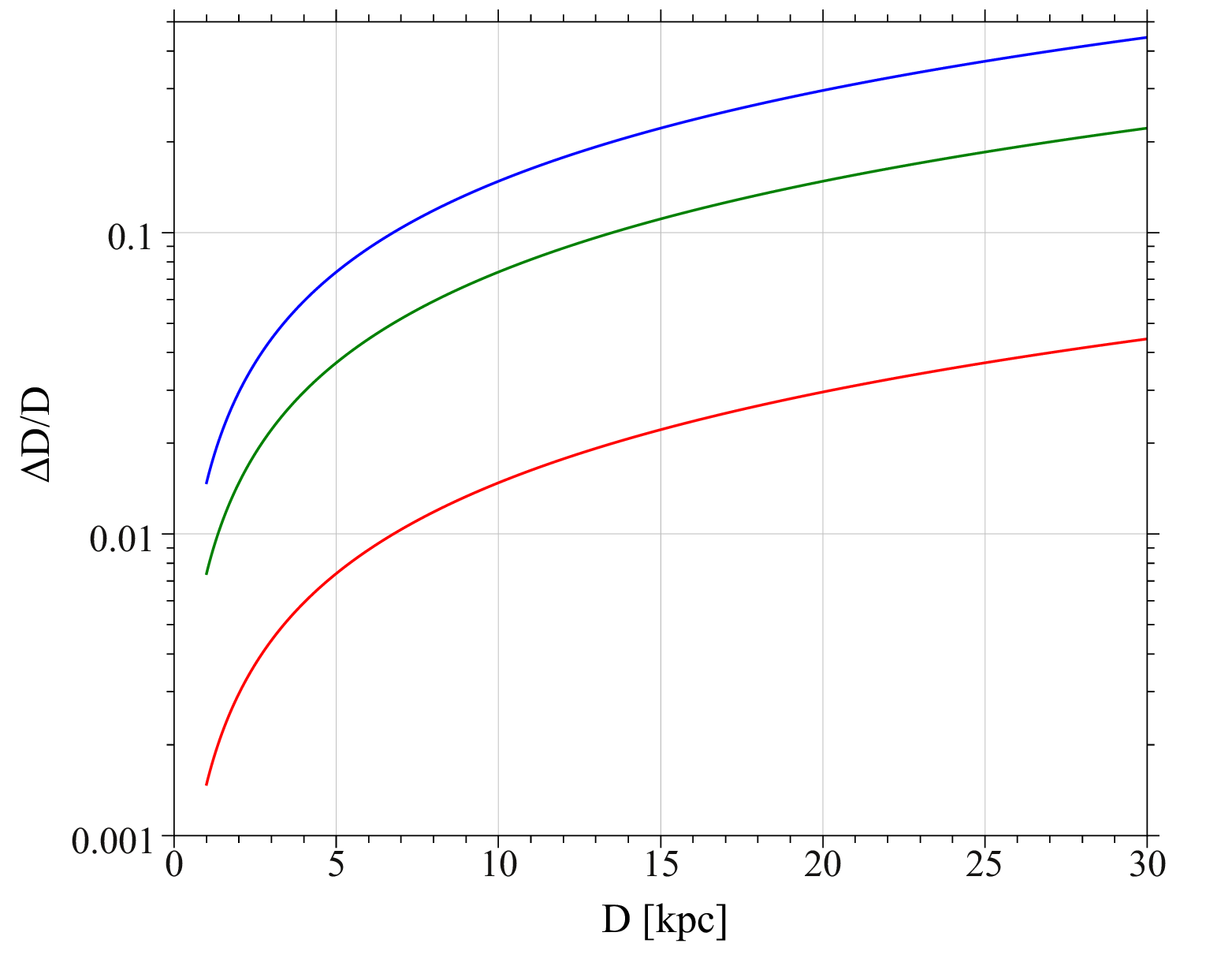,width=7cm}
\end{tabular}
}

\caption{Precision of parallax measurements from weekly  timing of MSP
  with the SKA.
{\em left)} Error in the timing parallax as a function of ecliptic
latitude. The top (blue) line assumes a time span of 5 years, the
lower (red) one
10 yr of observations. Both assume a timing precision of 10 ns.
{\em right)}  fractional precision of distance measurements for three
different timing precision values, 10 ns (red, bottom), 50 ns (green, middle) and
100ns (blue, top). See Smits et al. for details.
\label{fig:parallax}}
\end{figure}

Astrometric parameters can be determined in two ways for
pulsars. Firstly, using the telescope array as an imaging
interferometer, the pulsar can be treated as point source while
boosting the signal-to-noise ratio by gating the correlator to use
only signals during the few percent duty cycle when the pulsar is
actually visible. With images spread over a period of time, it will be
possible to measure parallaxes for nearly 10,000 pulsars with an
accuracy of 20\% or less \cite{stw+10}. Secondly, pulsar timing can
also be used for a precise determination of the position, proper
motion and parallax as all these parameters affect the arrival time of
the pulsars at our telescope on Earth. Here, MSPs with their higher
timing precision can be used more readily. Distances are retrieved via
a ``timing parallax'' which essentially measures the variation in
arrival time at different positions of the Earth is orbit due to the
curvature of the incoming wavefront. In contrast to an imaging
parallax, the sensitivity is highest for low ecliptic latitudes and
lowest for the ecliptic pole. Figure~\ref{fig:parallax} shows this
dependence of the parallax precision on ecliptic
latitude. Interestingly, it is still possible to measure a parallax
with finite precision at the ecliptic pole, in contrast to first
expectations. The origin of this is the small eccentricity of the
Earth orbit which allows us to still detect a variation of the arrival
time at different times of the year \cite{stw+10}. We expect that we
can measure distances of 20 kpc with a precision better than 20\% for
about 300 MSPs, while for some sources distances of 20 to 40 kpc can
be measured to 10\% or better.

A third method allows us to infer a distance to a binary MSP if we see
a relativistic decay of the orbit. The observed value for the orbital
period derivative will be altered from the intrinsic value by a
contribution arising from secular acceleration. As this depends on
distance, a comparison of the observed value with the one expected
from general relativity, allows us to determine the distance \cite{bb96}. In most
cases however, we need to apply the reverse method, i.e.~we need to
need to know the distance in order to derive the intrinsic orbital
decay rate, so that tests for theories of gravity can be performed.

\section{Tests of theories of gravity}

Binary pulsars already provide the best tests of theories of gravity
for strongly self-gravitating bodies \cite{ksm+06}. In the future,
with improved timing precision, we expect that these tests will even
surpass the precision of the current best tests of general relativity 
in the weak-field regime of the Solar System. In particular the
continued observations of the Double Pulsar will derive important
constraints for testing alternative theories of gravity and the
validity of specific concepts in strong gravitational fields, such as
the Lorentz-invariance \cite{wk07}. Most importantly, however, with
the SKA we will be able to probe the properties of black holes and
compare those to the prediction of general relativity for Kerr black holes. With
pulsars orbiting the super-massive black hole in the Galactic centre
and the discovery of binary pulsars with stellar-mass black hole
companions, we will be able to measure the mass, spin and quadrupole
moment of the black holes. These measurement will allow us to test the
{\em cosmic censorship conjecture} as well as the {\em no-hair
  theorem} \cite{kbc+04}.  The cosmic censorship conjecture states
that every astrophysical black hole, which is expected to rotate, has
an event horizon that prevents us from looking into the central
singularity. However, the event horizon disappears for a given value
of the black hole spin, so that we expect the measured spin to be
below the maximum allowed value. The no-hair theorem makes the
powerful statement that the black hole has lost all features of its
progenitor object, and that all black hole properties are determined
by only the mass and the spin (and possible charge). Therefore, if the
no-hair theorem is valid, the expected quadrupole moment of the black
hole can be unique determined from the mass and the spin. With a
measurement of all three quantities, this theorem can be tested
\cite{kbc+04}.  

With the chance to perform these experiments with the super-massive
black hole in the Galactic Centre, stellar black holes and, possibly,
intermediate mass black holes in globular clusters, a whole black hole
mass range can be studied. As already pointed out by Damour \&
Esposito-Farese \cite{de98}, a pulsar - black hole system will be  a
superb probe of gravity.

\section{Conclusions}

These technological advances will provide pulsar observers new tools
that will allow them to probe further into the unexplored regions of
the physical parameter space. The new telescopes combine a huge
increase in sensitivity with multi-beaming capability - a dream
combination for studies of pulsars and their exploitation as tools for
fundamental physics. The possible science applications extend far
beyond what has been, and can be, discussed here. A more complete
review can be found in the SKA science case \cite{ckl+04}, we will
however mention one other fascinating application in passing, as it
connects to the science drivers of the large optical telescopes: with
the SKA we will have the potential to find all the pulsars in the
Galaxy with radio beams pointed in our direction. The timing precision
in most cases should be sufficient to detect planets or minor bodies
possibly orbiting the neutron stars. Hence, for the first time, we
will not only get a census of radio emitting pulsars in the Galaxy but
also a census for planetary-sized objects around neutrons stars. Using
this statistical information we may finally have the chance to answer
the question: Why pulsar planets are so rare? Whatever the answer to
the formation process will be, it is clear that with the new
generation of radio telescope the future of pulsar astrophysics is
extremely bright.

%\bibliographystyle{ws-procs975x65}
%\bibliography{journals,psrrefs,psrrefs_add.bib,modrefs,crossrefs}

\end{document}